\documentclass[12pt]{article}
\usepackage{amsthm,amsmath}
\usepackage{mathtools}
\usepackage{amssymb}
\usepackage{verbatim}  
\usepackage{enumerate}
\usepackage{fancyvrb}
\usepackage{bm}
\usepackage{dsfont}
\usepackage{hyperref}
\usepackage{algorithm}
\usepackage{url} 
\RequirePackage{natbib}

\newcommand{\blind}{0}
\newcommand{\proc}{\mathbf{Y}(\mathbf{s})}
\newcommand{\covs}{\mathbf{X}}
\newcommand{\obs}{\mathbf{Z}(\mathbf{s})}
\newcommand{\nggt}{\boldsymbol{\epsilon}}
\newcommand{\eye}{\mathbf{I} }
\newcommand{\tree} { \mathbf{t} (\covs) }
\newcommand{\treehat} { \mathbf{ \hat t} (\covs) }
\newcommand{\tdsgn}{\mathbf{C}^k}
\newcommand{\tcntrst}{\boldsymbol{\pi}^k}
\newcommand{\esttcntrs}{\boldsymbol{\hat \pi}^k}
\newcommand{\forest}{\mathbf{f}({\covs})}
\newcommand{\spproc}{\boldsymbol{\nu}(\mathbf{s})}
\newcommand{\pars}{\boldsymbol{\theta}}
\newcommand{\estpars}{\boldsymbol{\hat \theta}}
\newcommand{\Sig}{\boldsymbol{\Sigma}}
\newcommand{\re}{\boldsymbol{\eta}}
\newcommand{\estre}{\boldsymbol{\hat \eta}}
\newcommand{\spatbas}{\mathbf{S}(\mathbf{s})}
\newcommand{\tspatbas}{\mathbf{S}^T(\mathbf{s})}
\newcommand{\proftree}{\mathbf{\hat t}(\covs | \Sig (\pars))}
\newcommand{\Om}{\boldsymbol{\Omega}}
\newcommand{\bbeta}{\boldsymbol{\beta}}
\newcommand{\ca}{\mathbf{c}^A}
\newcommand{\cat}{ \left(\mathbf{c}^A\right)^T}

\newcommand{\candset}{\mathbf{C}^A_i}

\begin{document}

\def\spacingset#1{\renewcommand{\baselinestretch}%
{#1}\small\normalsize} \spacingset{1}


\if0\blind
{
  \title{\bf Random Spatial Forests}
  \author{Travis Hee Wai\thanks{This publication was developed under a STAR research assistance agreements, No. RD831697 (MESA Air) and RD-83830001 (MESA Air Next Stage), awarded by the U.S Environmental Protection Agency. It has not been formally reviewed by the EPA. The views expressed in this document are solely those of the authors and the EPA does not endorse any products or commercial services mentioned in this publication.}  \\
    Department of Biostatistics, University of Washington\\ and\\
    	Michael T. Young\\
    Department of Environmental \& Occupational Health Sciences, \\University of Washington\\  and\\
	Adam A. Szpiro\\ 
	Department of Biostatistics, University of Washington
}
  \maketitle 
} \fi
\if1\blind
{
  \bigskip
  \bigskip
  \begin{center}
    {\LARGE\bf Random Spatial Forests}
\end{center}
	\medskip
} \fi

\begin{abstract}
We introduce random spatial forests, a method of bagging regression trees allowing for spatial correlation. Our main contribution is the development of a computationally efficient tree building algorithm which selects each split of the tree adjusting for spatial correlation. We evaluate two different approaches for estimation of random spatial forests, a pseudo-likelihood approach combining random forests with kriging and a non-parametric version for a general class of spatial smoothers. We show improved prediction accuracy of our method compared to existing two-step approaches combining random forests and kriging across a range of numerical simulations and demonstrate its performance on elemental carbon, organic carbon, silicon, and sulfur measurements across the continental United States from 2009-2010.
\end{abstract}

\noindent%
\vfill

\newpage
\spacingset{1.5} 
\section{Introduction}
\label{sec:intro}

Consider $n$ observations at locations $s_1,\ldots,s_n \in A \subset \mathbb{R}^d$ from the process
$$ \obs = \proc + \nggt, \quad \nggt \overset{\text{i.i.d}}{\sim} (0,\tau^2 \eye_n)$$
A fundamental problem in spatial statistics is reconstruction of the underlying process, $\proc$, on the basis of data that have independent measurement error, $\nggt$, incorporated. 
 Estimation of the underlying spatial process often focus on smoothing by kriging \citep{matheron1963principles} and it has become common to include spatially-indexed covariates  \citep{hengl2007regression}. Technological advancements have increased the ease and scope of data collection for these spatially-indexed covariates. For example, geographic information system (GIS) covariates from programs such as ArcGIS provide users with hundreds of covariates describing proximity variables to significant geographical features and buffer variables measuring geographic features within some radius. In addition, researchers have examined including additional sources such as satellite data \citep{xu2018national}, traffic data \citep{saucy2018land}, and meteorology data \citep{arain2007use}. 

A major purpose for estimation of the spatial process is to predict values at unobserved sites. Random forests \citep{breiman2001random} has been shown to be effective for prediction in high-dimensional scenarios and some have examined applying it to spatially-indexed covariates in order to estimate the underlying spatial process \citep{hu2017estimating,chen2018machine}. However, random forests does not incorporate information about the spatial locations of the data and some studies suggest that random forests and other machine learning methods to spatial data often do not yield any noticeable advantages over traditional geostatistical approaches such as kriging \citep{fox2018comparing,berrocal2019comparison}. In order to correct for this deficiency two-step approaches have been proposed where a spatial smoother is fit to the residuals from the random forests estimate. \cite{rolf2020post} provide a simple set of conditions under which this approach improves estimation accuracy and it has been shown to perform better than either using either method alone in practice \citep{liu2018improve}.  

Two-step optimization approaches are an inefficient optimization scheme.  Combining random forests and kriging can be viewed as an additive model,
$$\proc =  f(\covs) + \spproc,$$
with $f \left( \covs \right)$ a random forests estimate of the spatially-indexed covariates and a spatially correlated zero mean stochastic term, $\spproc \sim (\mathbf{0}, \Sig( \pars ))$, where the spatial covariance is known up to parameters $\pars$. In these applications the stochastic process is a statistical characterization of spatial variation not explained by the covariates, for example topography, climatological, and meteorological patterns, that are difficult to model explicitly. In order to maximize variability explained through the additive model, it is desirable to use random forests to model systematic variation which cannot be modeled in the spatial process. By ignoring the spatial correlation, random forests may model spatial structure that could have otherwise been included in the spatial process. Little has been done to explore the degree to which the predictive power of these models can be improved by incorporating spatial information into random forests itself. \cite{hengl2018random} proposed random forests for spatial data, where they explored adding geographic proximity as a covariate before applying the random forests algorithm but only found similar prediction accuracy to the two-step random forests kriging approach. We emphasize that our goal is to use random forests to utilize geographic covariates which model variation which could not be modeled by the spatial process, and including geographic proximity as a covariate does not help in that regard.

Our main contribution is a novel algorithm to construct spatially adjusted trees which allow for spatial correlation in sub-$\mathcal{O}(n^3)$ run time, and evaluate two different procedures for constructing random forests estimates from spatially adjusted trees. In section 2, we describe a summary of random forests and universal kriging, describe our modified tree building algorithm allowing for spatial correlation, and examine different approaches to constructing random forests estimates from spatially adjusted trees. In section 3, we provide simulation results demonstrating the advantage of our approach over two-step estimation strategies. In section 4, we apply our method to annual average elemental carbon (EC), organic carbon (OC), Silicon (Si), and Sulfur (S) across the continental United States for 2009-2010. We end the paper with a discussion of the advantages of our method and aspects for future work.

\section{Methods}
\label{sec:meth}

\subsection{Tree-Based Methods}

\subsubsection{Regression Trees}
\label{treelinmod}

Regression trees have gained popularity for their ability to approximate a wide variety of non-linear functions. Trees are built through an iterative process called recursive binary splitting, which aim to minimize \textit{tree impurity}, traditionally mean-squared error, through a greedy optimization approach. At each iteration, a new terminal node of the tree is created by an exhaustive search selecting the branch which minimizes tree impurity at the current step. Although trees are often described as segmenting the data into terminal nodes by following decision rules in an attempt to sort observations with similar covariates together, a regression tree can also be formulated as a linear model. 
 
A tree, $\tree$ with $k+1$ terminal nodes can be written as $\mathbf{ \tree}= \tdsgn \tcntrst $.  Each column $t$ of $\tdsgn$ is a vector indicating observations in the new terminal node created in iteration $t$, and its entries are

$$  C_{i t} = \begin{cases}
1, &  X_{j_t}(s_i) \leq r^t_{j_t} \text{and } i \text{ in terminal node being split}   \\
0 & \text{else} 
\end{cases} $$
with $X_{j_t}$ the covariate the splitting rule is created on, $r^t_{j_t}$ the selected cutpoint.

Similarly to a binary tree, each of the $k+1$ terminal nodes of the tree is encoded by a unique combination of the $k+1$ columns of $\tdsgn$ and the tree estimate for that terminal node is a unique linear combination of the corresponding entries of the $k+1$ vector $\tcntrst$.  
One particular advantage of treating a regression tree as a linear model is that by profiling out $\esttcntrs$, the total tree impurity depends only on the structure of the tree design matrix 
$$\|\obs - \mathbf{\tree} \|_2^2 = \obs^T \left(\eye_n - \tdsgn \left( (\tdsgn)^T \tdsgn \right)^{-1} (\tdsgn)^T \right) \obs,$$
leading to efficient computational methods since $\esttcntrs$ does not need to be optimized for every possible new branch.

\subsubsection{Random Forests}
\label{randforest}

While regression trees are able to approximate a wide variety of non-linear functions, they are often not good predictors alone due to their high variance. One method of variance reduction to improve prediction performance is bootstrap aggregation (bagging), an ensemble method of averaging over trees constructed on bootstrapped samples. The bagged estimate can be written as
$$ \mathbf{\hat f}({\covs}) = \frac{1}{B}\sum\limits_{i=1}^B \mathbf{t}^i(\covs^i)$$
with $B$ is the number of bootstrap replicates, $\mathbf{t}^i(\covs^i)$ the tree built on bootstrapped sample $i$. Optimal variance reduction occurs when each of the trees is independent, but in many cases trees built on bootstrapped sample tend to be similar. In order to minimize correlation between trees, random forests only uses a random subset of the covariates when creating a new terminal node for each tree. The process of bagging over trees in combination with the added randomization used in building a tree enables random forests to approximate a large class of functions while maintaining low generalization error.

\subsection{Spatial Statistics}

\subsubsection{Universal Kriging}
\label{univkrig}

Universal kriging is a widely used geostatistics method which incorporates spatial information available in the monitoring data with a linear function of the geographic covariates by adding a spatial correlation model. The universal kriging model can be structured as an additive model,
$$\proc = \covs \bbeta + \spproc$$ 
which contains a linear mean structure on the covariates and observations are subject to variation from the linear model by a realization of a spatial process $\spproc$. The kriging approach models the error term $\spproc$ as a realization of a Gaussian process and estimate $\pars, \bbeta$ by maximization of the log-likelihood.
\begin{align}
\underset{ \pars, \bbeta }{\text{argmax} } \, &-\frac{1}{2} \log | \Sig(\pars) | - \frac{1}{2} ( \proc - \covs \bbeta )^T \boldsymbol {\Sigma}^{-1} (\pars) (\proc- \covs \bbeta) \nonumber 
\end{align} 

For any fixed $\pars_0$, $ \bbeta$ which maximizes $\ell ( \bbeta,\pars_0 |\proc)$ is easily shown to be the generalized least squares estimator
$$ \boldsymbol {\hat \beta} = (\covs^T \Sig^{-1}(\pars_0) \covs )^{-1} \covs^T \Sig^{-1} (\pars) \proc.$$
The method of eliminating $\bbeta$ from the log likelihood by profiling is commonly employed, and universal kriging models are estimated by optimizing (\ref{proflik}). 
\begin{align}
\label{proflik}
\underset{ \pars}{\text{argmax} } \, &-\frac{1}{2} \log | \Sig(\pars) | - \frac{1}{2} (\proc - \covs \boldsymbol{ \hat \beta} )^T \Sig^{-1} (\pars) (\proc-\mathbf{ X} \boldsymbol{\hat\beta} ) \nonumber \\
& \text{s.t} \, \boldsymbol{ \hat \beta} = \left(\covs^T \Sig^{-1} (\pars)\mathbf{ X} \right)^{-1}\covs^T \Sig^{-1}(\pars) \proc
\end{align} 

\subsubsection{Efficient Estimation Strategies for Large Spatial Datasets}
\label{rankred}
Optimization of the log-likelihood in a universal kriging model involves inverting the covariance matrix, which is $\mathcal{O}(n^3)$. Recent work in making spatial statistics computationally feasible has relied on clever ways of structuring the covariance matrix to reduce the computational complexity in calculating its inverse \citep{banerjee2008gaussian,kaufman2008covariance}. Here, we take an approach following \cite{cressie2008} and decompose the spatial covariance matrix $\Sig (\sigma^2)= \sigma^2 \spatbas \tspatbas $.
This decomposition leads to the \textit{spatial mixed effects model} (\refeq{slmm}).
\begin{equation}
\obs = \covs\bbeta + \spatbas \re + \nggt \label{slmm}
\end{equation}
where $\spatbas \in \mathbb{R}^{n \times k}$ are \textit{spatial basis functions} and $\re \overset{\text{i.i.d.}}{\sim} (\mathbf{0},\sigma^2 \eye_k), \re \perp \nggt$ are the \textit{spatial random effects}. The class of basis functions which can be constructed by this procedure are detailed in section 3.1 of \cite{cressie2008} and are able to approximate covariance functions often used in spatial statistics \citep{nychka2002multiresolution}. 
Under the spatial mixed effects model, 
$$\obs \sim (\covs \bbeta , \sigma^2 \spatbas \tspatbas + \tau^2 \eye_n).$$ 
Predictions at unobserved locations follow from the  conditional expectation of the spatial mixed effects model given the realization of the spatial random effect 
$$\text{E}[\mathbf{Z(s_0)} | \estre ] = \covs_0 \boldsymbol{\hat \beta} + \mathbf{S(s_0)} \estre , $$
where $\boldsymbol{\hat \beta}$ is the best linear unbiased estimator, and $\estre$ is the best linear unbiased predictor.
The spatial random effect $\estre$ can be interpreted as a penalized regression estimator (\cite{ruppert2003semiparametric} 4.5.3). By Henderson's justification \citep{robinson1991blup}, optimizing $ \boldsymbol{\hat \beta}$ and $\estre$ leads to minimizing the criteria

$$ \|\proc- \covs \boldsymbol{ \beta} - \spatbas \boldsymbol{ \eta}\|_2^2 + \frac{\tau^2}{\sigma^2} \| \boldsymbol{\eta} \|_2^2,$$

which can be interpreted as ridge regression on $\estre$ with penalty $\lambda = \frac{\tau^2}{\sigma^2}$. 

\subsection{Spatially Adjusted Trees}

Additive models combining regression trees and kriging can be formulated as
\begin{equation}
\label{addmod}
\proc = \tree+ \spproc ,
\end{equation}
with $\tree$ the regression tree constructed from the covariates and $\spproc \sim N(0,\Sig (\pars))$ a realization of a Gaussian process. 

Under the additive model (\ref{addmod}), $\proc \sim N(\tree,\Sig(\pars))$. By maximum likelihood, we wish to find a regression tree estimate $\treehat$ and covariance parameters $ \estpars $ such that 
\begin{equation}
\label{pseudologlik}
\{ \treehat, \estpars \} = \underset{ \tree,\pars }{\text{argmax} }\, \left[ -\frac{1}{2} \log | \Sig(\pars) | - \frac{1}{2} (\proc - \tree)^T \Sig^{-1}(\pars) (\proc- \tree) \right]
\end{equation}

We propose a principled likelihood-based optimization motivated by profile likelihood. The regression tree is profiled out of the optimization problem as:
\begin{align}
\underset{\pars}{\text{argmax} } & \left[ -\frac{1}{2} \log | \Sig(\pars) | - \frac{1}{2} (\proc - \proftree)^T \Sig^{-1} (\pars) (\proc- \proftree) \right] \nonumber \\
& \text{s.t} \,\, \proftree = \underset{\mathbf{t} (\covs|\Sig(\pars)) }{\text{argmin} } \left(\proc- \mathbf{t} \left(\covs|\Sig(\pars) \right) \right)^T \Sig^{-1}(\pars) \left(\proc- \mathbf{t} \left(\covs|\Sig(\pars) \right) \right) \label{spatTreeLoss}
\end{align}

The dependence of the profiled spatially adjusted regression tree on the spatial correlation matrix is emphasized as $\proftree$. We note similarities of this optimization problem to universal kriging and traditional regression trees. In universal kriging, $\tree = \covs \bbeta$ and the profile likelihood optimization criteria selects $\bbeta$ which maximizes $\ell (\bbeta, \pars_0 |\proc)$ for some fixed $\pars_0$. On the other hand, if we ignore the spatial process and let $\Sig(\pars) = \eye_n$, there are no covariance parameters to maximize over and we would build a normal regression tree which minimizes mean squared error. Thus, a spatially adjusted regression tree should be built to minimize (\ref{spatTreeLoss}) for a given $\pars_0$. 

\subsubsection{Spatially Adjusted Tree Building Algorithm}
\label{spatTreesec}

We propose a novel, computationally feasible \textit{spatially adjusted tree building algorithm} to construct a spatially adjusted tree which aims to minimize (\ref{spatTreeLoss}) by recursive binary splitting. Note that this algorithm is a greedy approach and does not guarantee convergence to the true minimizer. In Section \ref{treelinmod}, we showed that each tree can be written as a linear combination of the tree design matrix $\mathbf{C}$ and their corresponding weights $\boldsymbol{\pi}$ so we can re-write \ref{spatTreeLoss} as:
\begin{equation}
\ell(\tdsgn,\tcntrst ) = \left(\proc- \tdsgn \tcntrst \right)^T \Sig^{-1} \left(\proc-\tdsgn \tcntrst \right) \label{treeloss}
\end{equation}
By profile likelihood, we define the "characteristic matrix" for the spatial tree building algorithm $\Om^k$ (\refeq{omeg}) which depends only on the tree design matrix.
\begin{equation}
\Om^k = \Sig^{-1} - \Sig^{-1} \tdsgn \left((\tdsgn)^T \Sig^{-1} \tdsgn \right)^{-1} (\tdsgn)^T \Sig^{-1} \label{omeg}
\end{equation}
The loss is solely a function of $\Om^k$, the characteristic matrix, and the observations $\proc$.  For any new branch, $\ca$, a vector noting which observations are in the new terrminal node, updating $\Om^{k+1}$ depends only  on $\ca$ and the previous characteristic matrix $\Om^k$ as
\begin{equation}
\Om^{k+1} = \Om^k - \Om^k \ca \left( \cat \Om^k \ca \right)^{-1} \cat \Om^k \label{omega}
\end{equation}
Details of equality for (\refeq{omega}) are included in the supplement. Using this fact, the change in loss between $\tdsgn$ and $\mathbf{C}^{k+1} = \begin{bmatrix} \tdsgn & \ca \end{bmatrix}$ is easily shown to be
\begin{equation}
\nabla \ell(\ca) = \ell(\tdsgn) - \ell(\mathbf{C}^{k+1}) = \proc^T \left(\Om^k \ca \left( \cat \Om^k \ca\right)^{-1} \cat \Om^k \right)\proc.
\end{equation}

The spatially adjusted tree building algorithm is summarized in Algorithm 1.

\begin{algorithm}
	\label{spatTree}
	\caption{Spatially Adjusted Tree Building Algorithm}
	\begin{enumerate}
		\item set $\mathbf{C}^0 = \begin{bmatrix} \mathbf{1}_n \end{bmatrix}$
		\item Given $\Sig$, set the initial value for $\Om^0= \Sig^{-1} - \mathbf{C}^0 \left((\mathbf{C}^0)^T \Sig^{-1} \mathbf{C}^0\right)^{-1} (\mathbf{C}^0)^T \Sig^{-1} $
		\item For $i = 1, 2, \ldots$
		\begin{enumerate}
			\item Take a random sample of the covariates $\covs_r \subset \{\covs_1, \covs_2, \ldots, \covs_p\}$
			\item Check each of the $i$ existing terminal nodes for a new terminal node created by a decision rule based on the sampled covariates $\covs_r$, to create a candidate set of possible splits $\candset$.
			\item Find the candidate split $\ca \in \candset$ which maximizes the change in loss 
			$$\proc^T \left(\Om^k \ca \left( \cat \Om^k \ca \right)^{-1} \cat \Om^k \right)\proc$$
			\item Update $$\Om^{k+1} = \Om^k - \Om^k \ca \left(\cat \Om^k \ca \right)^{-1} \cat \Om^k $$
			\item Repeat steps (a)-(c) until no more the maximum number of splits is exceeded or there are no more branches can be split without creating a branch with less than $m$ observations
		\end{enumerate}
	\end{enumerate}
\end{algorithm} 

\subsubsection{Computational Complexity for Spatially Adjusted Trees}
\label{compcmplx}

We first note that $\left( \cat \Om^k \ca \right)^{-1}$ is a scalar and the change in loss for any candidate split is
$$\nabla \ell(\ca) =  \frac{\left\| \cat\mathbf{w} \right\|_2^2}{\cat \Om^k \ca } =\frac{N(\ca)}{D(\ca)}, \quad \mathbf{w} = \Om^k \proc$$
Assume $\mathbf{X}_j(\mathbf{s})$ is unique at each location. For any covariate there are at most $n-1$ possible new cutoff values across all terminal nodes and these candidate splits are ordered. If the new candidate split is the first possible split of a terminal node, then it is a vector with a single one in position $l_i$ and zeros everywhere else. In this case, $\nabla \ell(\ca) = \frac{w^2_{l_i}}{\Omega^k_{l_i l_i}} $ and is $\mathcal{O}(1)$.  Otherwise, the next candidate split adds a single one in position $l_{i+1}$ to the previous split.  Then,
$$ \nabla \ell(\ca_{(i+1)j}) = \frac{ N (\ca_{ij}) + \mathbf{w}^2_{l_{i+1}}} { D (\ca_{ij}) + \Om^k_{l_il_i} + 2 \sum\limits_{m=1}^{i} \Om^k_{ l_m l_{i+1} }}.$$
The loss for each additional split can be computed in $\mathcal{O}(n)$ (compared to $\mathcal{O}(1)$ for standard regression trees) and the worst case run time for our spatially adjusted tree is $\mathcal{O}(pn^2 \log(n) +h)$, where $h$ is the run time to needed to invert the spatial covariance matrix a single time.  

\subsection{Random Spatial Forests: Pseudo-Likelihood Approach}
\label{spatRFPL}

Our algorithm describes a process for constructing spatially adjusted trees for known $\pars_0$, and we can construct spatial random forests estimates by aggregating over these trees. In practice however, $\pars$ is unknown. Since random forests estimates the expectation of an infinite tree (\cite{hastie2005elements}, 15.3.4), we propose a pseudo-likelihood approach where we replace the regression tree with its bagged random forests estimate in (\ref{spatTreeLoss}). 

Joint optimization of the random forests and the covariance parameters characterizing the spatial process is difficult for a number of reasons. For Matern covariance functions, the likelihood function is a non-convex function of the covariance parameters. Gradient based approaches for finding local minima/maxima cannot be applied for random forests since no closed form gradient exists. Further, numerical gradients are complicated by randomness in resampling of observations and covariates from random forests creating a stochastic function evaluation. 

In order to simplify estimation of the covariance parameters we approximate the spatial covariance as in section \ref{rankred}. Let 
$$\mathbf{V}(\kappa,\delta) = \kappa \mathbf{R}(\delta), \quad \mathbf{R}(\delta )= \left( \delta \spatbas \tspatbas + (1-\delta) \eye_n \right)$$ 
The profile log-likelihood can easily be shown to be written as a function of a single parameter $\delta$ by profiling out $\mathbf{\hat f}(\covs | \mathbf{R}(\delta))$ and $\hat \kappa$ as
\begin{align*}
\ell(\delta) = -\frac{n}{2}&\log \left( \hat \kappa \right) -\frac{1}{2}\left|\mathbf{R}(\delta) \right| - \frac{1}{2 \hat \kappa} \left(\proc - \mathbf{ \hat f}(\covs | \mathbf{R}(\delta) ) \right)^T \mathbf{R}^{-1}(\delta) \left(\proc -\mathbf{\hat f} ( \covs | \mathbf{V}(\hat\kappa, \delta) ) \right)\\
& \text{s.t} \,\, \mathbf{\hat f} (\covs|\mathbf{R}(\delta) ) = \underset{ \mathbf{f} (\covs|\mathbf{R}(\delta)) }{\text{argmin} } \left(\proc- \mathbf{f} \left(\covs|\mathbf{R}(\delta) \right) \right)^T \mathbf{R}^{-1}(\delta) \left(\proc- \mathbf{f} \left(\covs|\mathbf{R}(\delta) \right) \right) \\
& \text{and }\hat \kappa = \frac{1}{n} (\proc- \mathbf{\hat f}(\covs|\mathbf{R}(\delta)) )^T \mathbf{R}^{-1}(\delta) ( \proc- \mathbf{\hat f}(\covs|\mathbf{R}(\delta)) )
\end{align*}
This parameterization makes optimization simpler, as we only need to optimize over $\delta \in [0,1]$. Since we have a single parameter restricted to a small search space, we optimize the model by performing a grid search and selecting $\delta$ which minimizes the pseudo-likelihood $\ell(\delta)$.

\subsection{Random Spatial Forests: Non-Parametric Estimation}
\label{spatRFNP}
In the previous section, we derived an additive model using a pseudo-likelihood approach to integrate random forests into a likelihood model. However, it is not easy to interpret $\re$ as a random effect since it is difficult to imagine the data generating mechanism that might give rise to such fields \cite{hodges2016richly}. In this case, modeling the spatial process using a random effect is a form of regularization and pseudo-likelihood gives us a way to estimate the parameter $\delta$. The goal of modeling the air pollution surface in epidemiological studies is to produce accurate estimates for individuals at unobserved locations, which can be viewed as a prediction problem. An alternative criterion when prediction accuracy is desired, which is often the case in many statistical learning applications, is to minimize the expected mean squared test error:
$$\underset{\delta}{\text{argmin}} \, \text{E} \left[ \|\mathbf{Y} - \mathbf{\hat Y}(\mathbf{s},\delta)\|_2^2 \right].$$ 
Noting the relationship between $\delta$ and a penalty in section \ref{rankred}, a natural non-parametric approach would be to select the tuning parameter $\delta$ for the additive model by $k$-fold cross-validation in order to find $\delta$ which minimizes the out of sample test error. 

The addition of running $k$-fold cross-validation to estimate expected test error for each candidate $\delta$ results in a substantial increase in computational time. But a unique property of random forests is that since random forests only uses "out-of-bag" samples in its estimation, the resulting function $\mathbf{\hat Y} (\mathbf{s})$ is equivalent to its cross-validated estimate \cite{hastie2005elements}. This is desirable since the mean squared error of $ \mathbf{\hat Y} (\mathbf{s})$ on the training set is equivalent to its expected test error and makes $k$-fold cross-validation is unnecessary, reducing the computational burden. In order to leverage this property, we propose applying the random forests algorithm to aggregate our spatially adjusted trees and each their associated spatial smoothers as:
$$\mathbf{\hat Y}_t(\mathbf{s},\delta) = \frac{1}{B} \sum\limits_{i=1}^B \left[ \mathbf{\hat t}^i\left( \covs^i|\mathbf{R}^i(\delta) \right) + \mathbf{S}^i(\mathbf{s}) \estre^i \right] $$
For each bootstrap sample, we can use our tree building algorithm in section \ref{spatTree} to estimate both the spatially adjusted tree $\mathbf{t}^i\left( \covs^i|\mathbf{R}^i(\delta) \right)$ and its associated spatial random effect $\estre^i$ by its BLUP. Over a grid of $\delta \in [0,1]$, we fit $\mathbf{\hat Y}(\mathbf{s},\delta)$ using our spatially adjusted tree building algorithm and select $$\underset{\delta}{\text{argmin}} \,\|\obs-\mathbf{\hat Y}(\mathbf{s},\delta)\|_2^2.$$ 

\section{Simulation Study}

We conduct a set of simulations to compare different methods of combining random forests with a spatial smoother. Datasets for simulations are created on a grid of points over the continental United States spaced at 25km intervals and GIS covariates at these locations are provided from ArcGIS 10.2. 

\subsection{Generating the Observed Surface}
For each simulation, we constructed a fixed exposure surface from an additive model of a function of GIS covariates, $\forest$, and a fixed realization of a Gaussian process with exponential covariance process, $\spproc$, with range randomly generated between $10\% - 20\%$ of the maximum distance between points.  A variety of different generating functions were used for the GIS covariates, some which only used a small portion of the covariates, some which used all, and some which included interaction terms to induce non-linearity, but our simulations did not suggest the type of generating function had a significant impact on the results.

The observed surface is constructed as 
$$\proc = \gamma \forest + \spproc,$$
where the parameter $\gamma$ controls the proportion of variance attributable to the GIS covariates. We consider two scenarios for this parameter: (1) \textit{Strong Covariates}: $65\%$ of the generated process is due to the covariates (2) \textit{Weak Covariates}: $35\%$ of the generated process is due to the covariates.

\subsection{Methods combining Random Forests with Spatial Smoothing}
For our examples we formulate the spatial basis functions using TPRS following \cite{olives2014reduced}. This choice is arbitrary, and as noted in \ref{rankred} one could consider selecting alternative spatial basis functions. We selected TPRS as an alternative to kriging as there is an equivalence between thin plate regression splines (TPRS) and kriging with a Matern-class covariance with infinite range \citep{nychka2000spatial}. We compare six methods (1) Random Forests (RF)- implemented using the \texttt{randomForests} package (2) Spatial Smoothing (TPRS) - implemented by the \texttt{mgcv} package (3) Random Forests plus Spatial Smoothing (RF-TPRS): 2 step approach where first RF is run, then TPRS if fit to the RF residuals. (4) Spatial Smoothing plus Random Forests (TPRS - RF): 2 step approach where first TPRS is run, then RF is applied to the residuals from TPRS (5) Random Spatial Forests - Pseudo-Likelihood (SpatRF-PL), section \ref{spatRFPL} (6) Random Spatial Forests - Non-Parametric (SpatRF-NP), section \ref{spatRFNP}

\subsection{Evaluating Reconstruction Accuracy } 
\begin{figure}
	\includegraphics[height=0.5\textwidth] {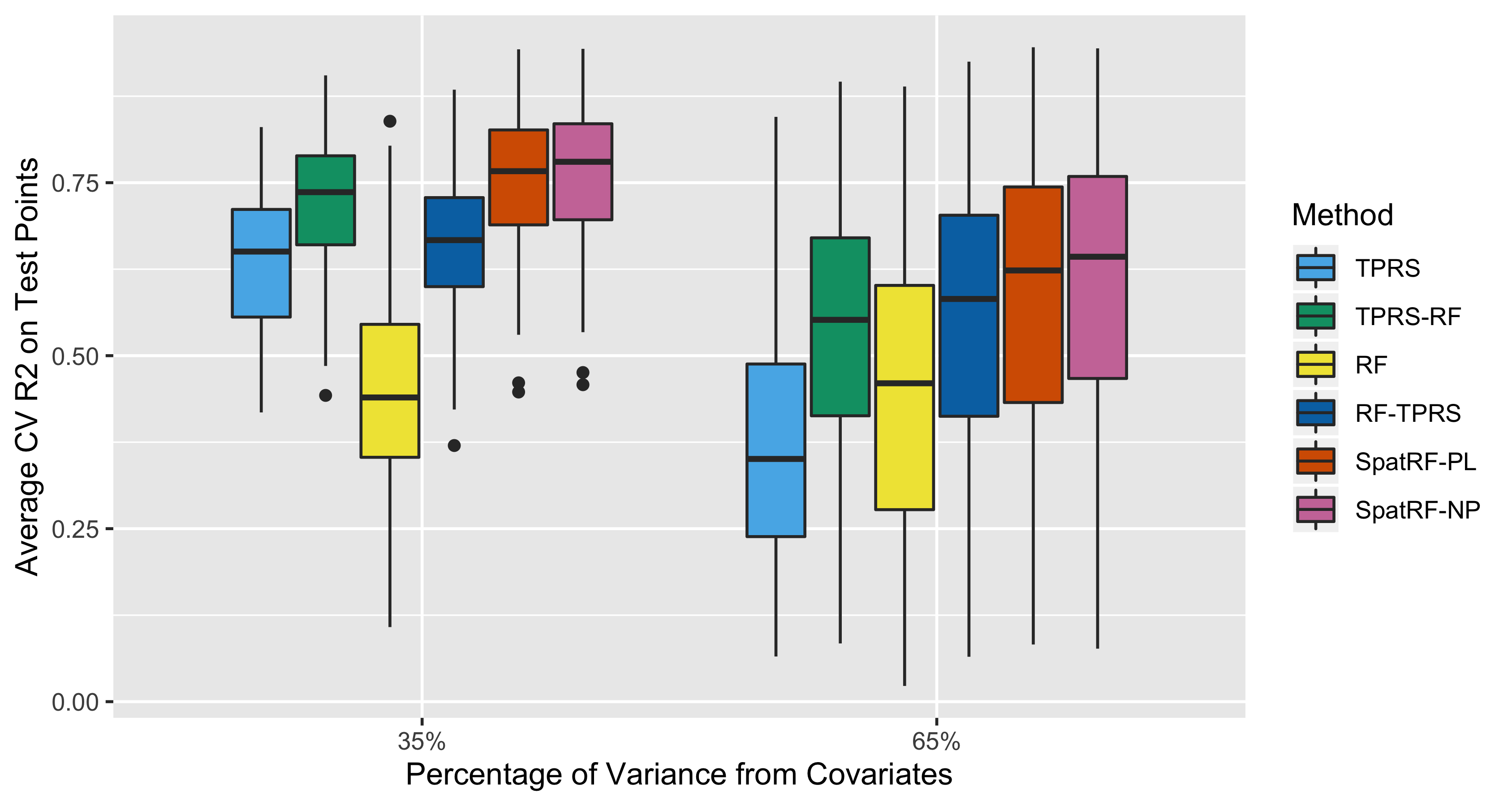}
	\caption{Simulation Results: Each point in the boxplot represents the average R\textsuperscript{2} at the 200 validation points for one of the 180 generated surfaces over 30 repeated samples. The box and whisker summarizes the prediction accuracy of the method on different simulated surfaces from a variety of generating functions. }
	\label{simresults}
\end{figure}

We generate a single observed surface $\proc$ and hold out 200 points for validation. For each simulation, 150 points are randomly sampled to train six different models to compare on. Training points are observed with independent measurement error 
$$ \mathbf{Z}(\mathbf{s}_\text{train}) = \mathbf{Y}(\mathbf{s}_\text{train}) + \nggt, \quad \nggt \sim N \left( \mathbf{0},\tau^2 \mathbf{I}_{n_\text{train}} \right),$$ 
where $\tau^2$ is randomly generated to be between 10\% and 25\% of the total variance. 
Model reconstruction accuracy is estimated by their prediction $R^2$ at validation points and we report the average $R^2$ on the validation points for each method from 30 different randomly sampled training points. This is repeated for 180 different observed surfaces.  Density plots of average $R^2$ for each method are shown in Figure \ref{simresults}. 

\subsection{Simulation Results}
In the \textit{strong covariates} scenario, RF does better than TPRS alone while this relationship is reversed in the \textit{weak covariates} scenario. This demonstrates that when a large percentage of the observed surface can be explained by the covariates, constructing a surface using a function of the covariates by RF performs better than ignoring the covariates and applying TPRS. On the other hand, when the covariates can only explain a small percentage of the total variation, using only the covariates via RF leads to worse prediction accuracy than simply applying TPRS alone. 

Additive models combining RF and TPRS (RF-TPRS and TPRS-RF) do better than either RF or TPRS alone.  Comparing RF-TPRS and TPRS-RF highlights the importance of the optimization approach. Although RF-TPRS and TPRS-RF are both composed of a random forests and thin plate regression spline, the order of estimation can have a large impact on the models prediction accuracy. When the covariates are responsible for a large percentage of variability in the observed surface RF-TPRS performs noticeably better than TPRS-RF, and vice versa when the covariates explain a small portion of the variance. Our methods demonstrate how constructing random forests allowing for spatial correlation leads to more accurate predictions than either two-step approach regardless of how much variability can be explained by the covariates. In our simulations, SpatRF-PL and SpatRF-NP \label{key}have better prediction accuracy than RF-TPRS and TPRS-RF in all scenarios. Comparing our two methods, SpatRF-NP performs slightly better than SpatRF-PL in both cases. 

\section{Application to Sub-Species of PM\textsubscript{2.5}}

We develop air pollution models for annual averages of four PM\textsubscript{2.5} (particulate matter less than 2.5 $\mu$m in aerodynamic diameter) sub-species: elemental carbon (EC), organic carbon (OC), silicon (Si), and sulfur (S) using Environmental Protection Agency (EPA) Interagency Monitoring for Protected Visual Environments (IMPROVE) and Chemical Speciation Network (CSN) monitoring data from 2009-2010. Following \cite{bergen2013national}, we only include CSN and IMPROVE monitors with at least 10 data points per quarter and no more than 45 days between consecutive measurements. Si and S measurements were averaged over 01/01/2009–12/31/2009, while EC/OC consisted of measurements from 204 IMPROVE and CSN monitors averaged over 01/01/2009–12/31/2009, and measurements from 51 CSN monitors averaged over 05/01/2009 - 04/30/2010. Annual averages were square-root transformed prior to modeling. 

In addition to methods used in the simulations, we include universal kriging estimates which deal with the high dimensionality of the covariates by pre-processing the covariates by partial least squares (UK-PLS) and use an exponential covariance matrix. This technique was employed in the original analysis by \cite{bergen2013national} and is commonly employed in many land-use regression settings. The reported $R^2$ values for UK-PLS are not identical to those reported in by \cite{bergen2013national} since they calculated $R^2$ on the square root scale, while we transform predictions back to the original scale before computing $R^2$ values. Additionally, we examine random forests with spatial information included (RF w/ TPRS). This approach included spatial basis functions are included as covariates accounting for geographic proximity between observations and is a heuristic for including spatial information into normal random forests. 

Surface reconstruction accuracy of each methods is assessed by comparing predictions generated from ten-fold cross-validation. Performance of each model is based on their average cross-validated $R^2$ over ten separate cross-validation runs in Table \ref{cvr2}. Cross-validated prediction accuracy over the different components of PM2.5 show similar findings to our simulation results, with SpatRF-NP consistently performing at least as well as any of the alternative methods. Figures showing predicted annual averages of EC, OC, S, and Si across the continental United States are included in the supplement.

\begin{table}[]
	\begin{tabular}{|c|c|c|c|c|c|c|}
		\hline
		& UK-PLS & RF w/ TPRS & RF-TPRS & TPRS-RF & SpatRF - PL & SpatRF - NP \\ \hline
		EC & 0.73 & 0.80     & 0.81   & 0.71   & 0.83       & 0.83     \\ \hline
		OC & 0.58  & 0.60      & 0.59   & 0.57   & 0.59      & 0.63       \\ \hline
		Si & 0.56  & 0.57      & 0.53   & 0.59   & 0.58       & 0.60       \\ \hline
		S  & 0.95  & 0.93      & 0.89   & 0.94   & 0.94       & 0.94       \\ \hline
	\end{tabular}
	\caption{Ten-fold cross-validated prediction accuracy, summarized by $R^2$, of each method for PM\textsubscript{2.5} components Elemental Carbon (EC), and Organic Carbon (OC), Silicon (Si), Sulfur (S) collected by AQS and IMPROVE monitoring networks from 2009-2010. }
	\label{cvr2}
\end{table}

When a large proportion of the variance can be explained by the covariates, demonstrated by EC and OC, RF performs better than TPRS and RF-TPRS has improved cross-validated accuracy over TPRS-RF. SpatRF-NP and SpatRF-PL show small but noticeable improvements over RF-TPRS and are more accurate for both pollutants. In these examples using random forests instead of using a linear model with dimension reduction on the covariates by PLS can yield noticeable improvements in prediction accuracy as SpatRF-PL and SpatRF-NP show noticeable increases in cross-validated $R^2$ over UK-PLS. 

Si and S are examples where spatial smoothing is able to model a larger proportion of the variance then the covariates alone. In both of these cases, RF performs worse $R^2$ than TPRS and RF-TPRS has lower cross-validated accuracy TPRS-RF. Interestingly, TPRS alone has better cross-validated accuracy than RF-TPRS suggesting that applying spatial smoothing to the residuals from random forests does not guarantee that the combined approach is more accurate than either individual method alone. For S, TPRS-RF, UK-PLS, SpatRF-PL, and SpatRF-NP all do quite well (CV $R^2$ 0.94-0.95), while SpatRF-NP has the highest cross-validated $R^2$ for Si.


\section{Discussion}

This paper presents a novel interpretation of regression trees in the form of a linear model, suggesting a principled approach to estimating regression trees which allow for correlation. By carefully constructing the tree design matrix, we show that this approach lends itself to efficient computation by taking advantage of its block structure. Through simulation results and on observed annual average EC, OC, Si, and S from 2009-2010, we demonstrate that this approach results in more accurate predictions than two-step estimation methods. 

In this paper we examined a random forests algorithm using our novel tree building algorithm to adjust for spatial correlation. Another popular tree based ensemble method is boosting, and it would be straightforward to apply our tree building algorithm to boost spatially adjusted trees. Our tree building algorithm adjusting for correlation is also not restricted to estimation in spatial applications. For example, prediction problems where it is desirable to adjust for correlation occurs in other application such as network-linked data \cite{li2019prediction}. 

The general approach of formulating a tree as a linear model would suggest that we could extend this method to generalized linear models (GLM) by adjusting the tree impurity metric to the negative log-likelihood of the selected GLM. However, this approach is computationally difficult. For the identity link, parameter estimates for the contrast vector $\boldsymbol{\pi}$ are profiled out, leading to a search only over candidate split vectors. For GLMs, there are no general closed form estimates for the corresponding parameters, thus each candidate split would require an inner optimization to obtain estimates for $\boldsymbol{\pi}$, which we suspect would make this approach prohibitively computationally intensive.

Our optimization approach took advantage of recent computational developments in spatial statistics to reduce the parameterization of the covariance to a single parameter $\delta$ and selecting its value by grid search. We note here that this is not required, for example, one could consider using a normal kriging covariance with an exponential covariance function and select the parameters by grid search, but adding additional parameters becomes computationally expensive since the number of points to consider scales exponentially. Bayesian optimization and covariance matrix adaptation - evolution strategy have been used in the machine learning literature for gradient free optimization of "black-box" prediction models with stochastic function evaluation where multiple tuning parameters need to be selected. Both of these methods can be applied to random spatial forests and are easily parallelizable to make optimization feasible for more complex covariance function parameterizations.

\bibliography{./spatrf_arxivv2}

\end{document}